\documentstyle[inputenc,prl,aps,graphicx,amsmath,preprint]{revtex}

\inputencoding{ansinew}

\begin{document}
\draft
\title{{\bf Exchange instability of the two-dimensional electron gas in
semiconductor quantum wells}}
\author{A.R.~Go\~ni$^1$, U.~Haboeck$^1$, C.~Thomsen$^1$, K.~Eberl$^2$, F.A.~Reboredo$%
^3$, C.R.~Proetto$^3$, and F.~Guinea$^4$}
\address{$^1$Institut f\"{u}r Festk\"{o}rperphysik, Technische\\
Universit\"at Berlin, Hardenbergstr. 36, 10623 Berlin, Germany}
\address{$^2$MPI f\"ur Festk\"orperforschung, Heisenbergstr. 1,\\
70569 Stuttgart, Germany}
\address{$^3$Comisi\'on Nacional de Energ\'{\i}a At\'omica, Centro At\'omico\\
Bariloche, 8400 S.C.~de Bariloche, Argentina}
\address{$^4$Instituto de Ciencias de Materiales (CSIC), Cantoblanco,\\
28049 Madrid, Spain}
\maketitle

\begin{abstract}
\noindent {\small {\em A two-dimensional (2D) electron gas formed in a
modulation-doped GaAs/AlGaAs single quantum well undergoes a first-order
transition when the first excited subband is occupied with electrons, as the
Fermi level is tuned into resonance with the excited subband by applying a
dc voltage. Direct evidence for this effect is obtained from low-temperature
photoluminescence spectra which display the sudden renormalization of the
intersubband energy $E_{01}$ upon the abrupt occupation of the first excited
subband. Calculations within density-functional theory, which treat the 2D
exchange potential {\it exactly}, show that this thermodynamical instability
of the electron system is mainly driven by {\it intersubband} terms of the
exchange Coulomb interaction. From temperature-dependent measurements the
existence of a critical point at $T_c = 35\pm 5$ K is inferred.}}
\end{abstract}

\pacs{PACS: 71.30.+h, 78.55.-m, 71.10.-w, 73.21.Fg}

\preprint{} \sloppy




{\normalsize 
}

{\normalsize Exchange effects are a fundamental manifestation of
electron-electron interactions in many-particle systems like high-mobility
two-dimensional (2D) electron gases which form in modulation-doped
semiconductor quantum well structures. Exchange-correlation interactions are
at the origin of a vast variety of fascinating quantum phenomena in solids
\cite{mahan81a}, for example, spin excitations, magnetic ordering, excitonic
binding and band-gap renormalization among others. The latter, for instance,
is apparent in semiconductor heterostructures as a reduction of gap energies
when the corresponding band states become populated with carriers either by
varying doping levels \cite{delal87a,gonix99a} or under intense
photoexcitation \cite{cingo90a}. This effect is explained as arising from
exchange-correlation corrections due to the presence of free carriers in the
system \cite{klein85a,dassa90a}. }

{\normalsize Electronic correlations dictate the behavior of dilute electron
gases of reduced dimensions particularly at very low densities. For instance,
the occurrence of a metal-insulator transition (MIT) in two dimensions at zero
magnetic field \cite{kravc94a,ernst94a} is in open contrast with the prediction
of scaling theory that a {\it noninteracting} 2D electron or hole system becomes
localized at low temperatures for any degree of disorder \cite{abrah79a}.
Recently, other thermodynamic properties of 2D electron gases like the chemical
potential and its density derivative, the compressibility, have been found to
exhibit anomalous behavior across the metal-insulator transition deviating from
what is expected within Hartree-Fock theory \cite{ilani99a}. In spite of the
universalities detected at the MIT, for example in the temperature dependence of
the resistivity \cite{kravc95a}, this transition does not reveal a purely
intrinsic property of interactions in low-dimensional systems since disorder
plays the crucial role. On the contrary, localization effects can be widely
suppressed in double-layer electron gases, in which the electron gas in one
quantum well is used for the complete screening of the disorder potential of the
ionized donors \cite{eisen92a}. Using a capacitance technique Eisenstein {\it et
al.} \cite{eisen92a} were able to show that at low densities but without
magnetic field the compressibility of a high-mobility 2D electron gas becomes
{\it negative} owing mainly to exchange interactions. This leads to a
thermodynamical instability of the electron system. Furthermore, exchange
effects might induce other types of instabilities in double quantum wells such
as a (controversial) bilayer-to-monolayer transition \cite{zheng97a} or even one
towards a magnetic ground state \cite{rebor99a}. The exchange interaction also
gives rise to first-order transitions in diluted magnetic semiconductors
\cite{breyx00a}. }

{\normalsize In this Letter we show that at low temperature and zero magnetic
field a 2D electron gas formed in a single GaAs quantum well undergoes a
first-order phase transition, as the first excited electron subband becomes
populated with electrons by raising the Fermi level with a gate voltage. The
evidence is found in the sudden and abrupt renormalization of the energy of the
first excited subband, as determined from photoluminescence and inelastic light
scattering measurements. Self-consistent density-functional calculations with
{\it exact} exchange potential for a 2D electron system reveal that this
transition is driven by intersubband exchange interactions which provide a
feedback mechanism for charge transfer into the excited subband. Experiments
performed at different temperatures indicate the existence of a critical point
for this transition at around $35\pm 5$ K. }

{\normalsize 
}

{\normalsize The sample consists of a modulation-doped 245 \AA ~-wide GaAs
single quantum well (SQW) with Al$_{0.33}$Ga$_{0.67}$As barriers grown by
molecular-beam epitaxy. The growth sequence is given elsewhere \cite {ernst94a}.
Without bias only the lowest subband is occupied with electrons with Fermi
energy $E_{F}\approx $~25 meV. The energy separation to the second subband is
$E_{01}\approx $~28 meV. The electron gas is contacted from the surface by In
alloying in order to apply a dc bias up to 30 V between it and a metallic back
contact. Photoluminescence (PL) and inelastic light scattering spectra were
excited with a tunable Ti:sapphire laser and recorded with optical multichannel
detection.}

{\normalsize 
}

{\normalsize The effect of the applied voltage is an increase of the
electron density in the quantum well shifting the Fermi level in the
structure towards degeneracy with the bottom of the first excited subband.
Figure 1 shows some PL spectra recorded at 7 K and for different bias. The
low-energy emission line $E_{0}$ corresponds to recombination processes
between the lowest electron and heavy-hole subbands (with subband index 0).
The peak labeled as $E_{1}$ is associated with optical transitions between
the first excited electron subband and the hole ground state, which become
dipole allowed due to the lack of inversion symmetry of the triangular
potential in the doped well. The peak at 1.515 eV originates from near
bandgap emission of the GaAs buffer layer. At high voltages a high-energy
cutoff at the Fermi energy $E_{F}^{\prime }=E_{F}(1+m_{e}/m_{h})$ is clearly
apparent from the PL spectra. The factor containing the ratio of electron
and hole effective masses accounts for the curvature of the valence band.
With increasing bias the Fermi energy increases leading to a population of
the first excited subband. An important result concerns the strong redshift
of the subband energy $E_{1}$ by about 11 meV. Moreover, the occupation of
the second subband proceeds abruptly at the voltage for which the Fermi
level reaches its bottom. This reduction of $E_{1}$ is a consequence of
bandgap renormalization effects due to exchange-correlation corrections in
the presence of free carriers \cite{klein85a,dassa90a}. }

{\normalsize The dependence on bias of the electron densities $n_{0}$ and $%
n_{1}$ of the ground state and first excited subband, respectively, as well as
the intersubband energy $E_{01}$ has been determined from a quantitative
analysis of PL line shapes, as described elsewhere \cite{gonix99a,ernst94a}. The
values for $n_{0}$, $n_{1}$ and $E_{01}$ obtained at 10 K are plotted in Fig. 2
as a function of the Fermi level referred to the top of the valence band. Below 40 K when
the Fermi level equals $E_{1}$ the electron density $n_{1}$ jumps from zero to a
finite value ranging from 3 to $8\times 10^{10}$ cm$^{-2}$ depending on
temperature. The electron density of the lowest subband, in contrast,
increases slightly but smoothly with voltage. Simultaneously with the abrupt
filling of the second subband, a sudden reduction of the intersubband energy
$E_{01}$ by about 3.5 meV is observed \cite{notex01a}. Bandgap renormalization
acts as a feedback mechanism for subband filling leading to a sudden population
of the excited electron subband and, in addition, causing the pronounced
decrease of the intersubband spacing in the well. We point out that PL provides
a means for the determination of the Fermi level which is independent of the
quality of the contacts or the way in which charge flows into the quantum well.
We thus rule out spurious charging effects as cause of the observed jumps in 2D
density and intersubband spacing. }

{\normalsize The observed discontinuities in electron density and gap
renormalization upon occupation of the first excited subband point to a
thermodynamical instability of the 2D electron gas. In order to enlighten this
fundamental issue, we have performed self-consistent calculations of the subband
structure and level occupation of the single quantum well within the formalism
of density functional theory (DFT). Our calculation differs from the standard
local-density approximation (LDA) in two mayor points. The exchange interaction
is treated {\it exactly}\cite{stadele} for a quasi-2D electron system and the
number of particles is {\it not fixed} but is allowed to change. Mathematically,
the exchange potential is calculated according to
\begin{eqnarray}
V_{x}({\bf r}) &=&\frac{\partial E_{x}}{\partial \rho \left( {\bf r}\right) }%
=\int {\bf dr}^{\prime }\left\{ \sum_{vk}\int {\bf dr}^{\prime \prime }\left[
\frac{\partial E_{x}}{\partial \phi _{n{\bf k}}\left( {\bf r}^{\prime
}\right) }\frac{\partial \phi _{n{\bf k}}\left( {\bf r}^{\prime }\right) }{%
\partial V_{KS}\left( {\bf r}^{\prime \prime }\right) }+c.c.\right] \right. +
\label{eq:exact1} \\
&&+\left. \sum_{v}\left[ \frac{\partial E_{x}}{\partial k_{F}^{v}}\frac{%
\partial k_{F}^{v}}{\partial V_{KS}\left( {\bf r}^{\prime \prime }\right) }%
\right] \right\} \frac{\partial V_{KS}\left( {\bf r}^{\prime \prime }\right)
}{\partial \rho \left( {\bf r}\right) },	\nonumber
\end{eqnarray}
where $E_{x}$ is the exact exchange energy, $\rho \left( {\bf r}\right) $ is
the electron density, $V_{KS}({\bf r})=V_{ex}({\bf r})+V_{H}({\bf r})+V_{x}(%
{\bf r})+V_{c}({\bf r})$ is the Kohn-Sham potential given as sum of the
external, Hartree, exchange and correlation potentials, respectively, and $%
\phi _{n{\bf k}}\left( {\bf r}\right) \propto e^{i{\bf k.\rho }}\cdot \xi
_{n}\left( z\right) $ is the wave function of an electron in the quantum
well characterized by the envelope function $\xi _{n}\left( z\right) $ in the
direction of confinement $z$. The fist term in Eq. (\ref{eq:exact1}) represents
derivatives of $E_{x}$ with respect to the {\it shape} of the wave
functions\cite{stadele}, whereas the second one accounts for the variation of
the exchange energy with the occupations. }

{\normalsize Integration in the $(x,y)$ plane yields (in atomic units)
\begin{eqnarray}
V_{x}\left( z\right) &=&\frac{1}{2}\left\{ \sum_{vv^{\prime }n}I_{1}\left(
v,v^{\prime },n\right) \int dz^{\prime }\frac{\xi _{v}\left( z^{\prime
}\right) \xi _{n}\left( z^{\prime }\right) }{\varepsilon _{v}-\varepsilon
_{n}}\chi ^{-1}\left( z,z^{\prime }\right) +\right.  \label{eq:fullpot} \\
&&\left. +\sum_{vv^{\prime }}I_{2}\left( v,v^{\prime }\right) \int
dz\,^{\prime }\,\left| \xi _{v}\left( z^{\prime }\right) \right| ^{2}\chi
^{-1}\left( z,z^{\prime }\right) \right\} ,	\nonumber
\end{eqnarray}
\begin{equation}
I_{1}\left( v,v^{\prime },n\right) =k_{F}^{v}k_{F}^{v^{\prime }}\int \frac{%
d\rho }{\rho }J_{1}\left( k_{F}^{v}\,\rho \right) J_{1}\left(
k_{F}^{v^{\prime }}\,\rho \right) \int dz_{1}dz_{2}\frac{\xi _{v}\left(
z_{1}\right) \xi _{n}\left( z_{2}\right) \xi _{v^{\prime }}\left(
z_{1}\right) \xi _{v^{\prime }}\left( z_{2}\right) }{\sqrt{\rho ^{2}+\left(
z_{1}-z_{2}\right) ^{2}}},	 \label{eq:I1}
\end{equation}
\begin{equation}
I_{2}\left( v,v^{\prime }\right) =k_{F}^{v^{\prime }}\int d\rho J_{0}\left(
k_{F}^{v}\,\rho \right) J_{1}\left( k_{F}^{v^{\prime }}\,\rho \right) \int
dz_{1}dz_{2}\frac{\xi _{v}\left( z_{1}\right) \xi _{v}\left( z_{2}\right)
\xi _{v^{\prime }}\left( z_{1}\right) \xi _{v^{\prime }}\left( z_{2}\right)
}{\sqrt{\rho ^{2}+\left( z_{1}-z_{2}\right) ^{2}}},  \label{eq:I2}
\end{equation}
where the operator $\chi \left( z,z^{\prime }\right) =\partial \rho \left(
z\right) /\partial V_{KS}\left( z^{\prime }\right) $ is related to the
Linhard susceptibility of the 2D gas in the T}$\longrightarrow ${\normalsize %
0 limit. In Eq. (\ref{eq:fullpot}), $v\,$ and $v^{\prime }$ sum only
occupied subbands, while the index $n$ runs over all subbands (with
exception of $n=v$). $J_{i}(z)$ stands for the Bessel function of order $i$,
$\rho $ is the in-plane coordinate vector and $k_{F}^{v}$\thinspace \ and $%
k_{F}^{v^{\prime }}\,$\ are the Fermi wave vectors of the occupied subbands.
}

{\normalsize The striking and novel result of our theory is that the exchange
potential $V_{x}(z)$ changes discontinuously every time a subband becomes
occupied with electrons. For our experimental situation, i.e. $v^{\prime }=0$
and $v=1$, the second term in Eq. (\ref{eq:fullpot}) is finite even for $%
k_{F}^{v=1}\longrightarrow 0^{+}$ because $I_{2}\left( 1,0\right) \neq 0$,
whereas the term containing $I_{1}\left( 1,0,n\right) $ goes to zero. The
operator }$\chi
^{-1}\left( z,z^{\prime }\right) ${\normalsize \ also presents a discontinuity
as soon as $k_{F}^{v=1}>0$. The self-consistent solution of the Kohn-Sham
equations with }$V_{x}\left( z\right) $ {\normalsize given by Eq.
(\ref{eq:fullpot}), in fact, exhibits an abrupt jump from a phase with one
subband occupied to a phase with two occupied subbands when the Fermi level
equals $ E_1$. We obtain at occupation of subband 1 a sudden reduction of
$E_{01}$ by about 2.5 meV for an initial value of 26.5 meV, which is in good
quantitative agreement with the experimental observation. This is a clear
signature of an exchange-driven first-order phase transition. This drop of the
intersubband spacing is due to an effective {\it level repulsion} between the
first excited subband and the Fermi energy due to exchange interaction. Due to
intersubband exchange terms it is more favorable for the 2D electron gas to
transfer a macroscopic amount of charge into the first excited subband upon
occupation, thus, leading to the sudden collapse of the subband spacing
$E_{01}$. Details of the theory and the self-consistent calculations will be
given elsewhere\cite{exacttheory}.}

{\normalsize Further evidence for the first-order character of the exchange
instability in electron gases by occupation of higher subbands can be gained
from PL measurements at different temperatures. Figure 3 displays several
isothermal curves representing the variation of the Fermi level and the
density of the first excited subband. The isotherms fall into two classes
according as the temperature lies below or above a critical temperature $T_c
= 35\pm 5$ K. Below $T_c$ the 2D electron gas can exist in two states with
and without population of the first excited subband; its occupation proceeds
abruptly due to the exchange instability. Above the critical temperature, on the
contrary, the discontinuity in the density $n_1$ disappears and the system
evolves continuously from one phase to another. In addition, evidence of a phase
mix revealing a spatial inhomogeneity of the electron gas is obtained from
luminescence. At the instability and with fixed voltage PL spectra display a
multiple-peak time-varying structure at energies around $E_1$. Each peak of this
feature is assigned to PL emission at $E_1$ arising from different regions
within the laser spot with slightly different electron density.}

{\normalsize 
}

{\normalsize In conclusion, we have found that a 2D electron gas formed in a
modulation-doped GaAs single quantum well undergoes a first-order phase
transition when the first excited subband becomes populated by raising the Fermi
level in the structure using a dc voltage. The signature of this transition is
seen in the abrupt renormalization of the subband energy and the associated jump
in electron density upon occupation of the excited subband. Furthermore, we have
determined from the isotherms a critical temperature of about 35 K for the
transition from the phase without to that with finite occupation of the excited
subband. Self-consistent calculations within density-functional theory, which
treat the exchange interaction in the 2D system exactly, show that such
thermodynamical instability of the electron gas is mainly induced by
intersubband exchange terms of the Coulomb interaction. This instability
corresponds, for instance, to the same universal class of liquid-vapor phase
transitions. In spite of the importance of thermodynamical fluctuations in two
dimensions, they are not sufficiently large in order to prevent the
exchange-driven transition from being of first order. In this way, we have
provided further insight into the issue of the many-body behavior of
high-mobility electron gases in 2D arising from exchange interactions between
correlated electrons in different subbands of semiconductor quantum wells. }

{\normalsize This work is supported in part by the Deutsche
Forschungsgemeinschaft in the framework of Sfb 296. F. A. R. and C. R. P.
acknowledge support from CONICET of Argentina. F.G. acknowledges financial
support from CICyT (Spain), through grant PB96-0875.}


\def\fn{fig1ins}
\begin{figure}[tbp]
\vspace*{1ex}
\caption{Photoluminescence spectra of a modulation-doped 25 nm wide single
quantum well at 7 K and for different gate voltages. The position of
bandgaps and Fermi energy is indicated. The peak at 1.515 eV arises from
bulk GaAs luminescence. }
\label{PLbias}
\end{figure}

\def\fn{fig2ins}
\begin{figure}[tbp]
\vspace*{1ex}
\caption{Dependence on gate voltage of {\bf (a)} the electron densities $n_0$
and $n_1$ of the lowest and the first excited subband, respectively, and
{\bf (b)} of the intersubband spacing $E_{01}$, as determined from PL
spectra at 10 K. The curves are a guide to the eye. }
\label{2Dparam}
\end{figure}

\def\fn{fig3ins}
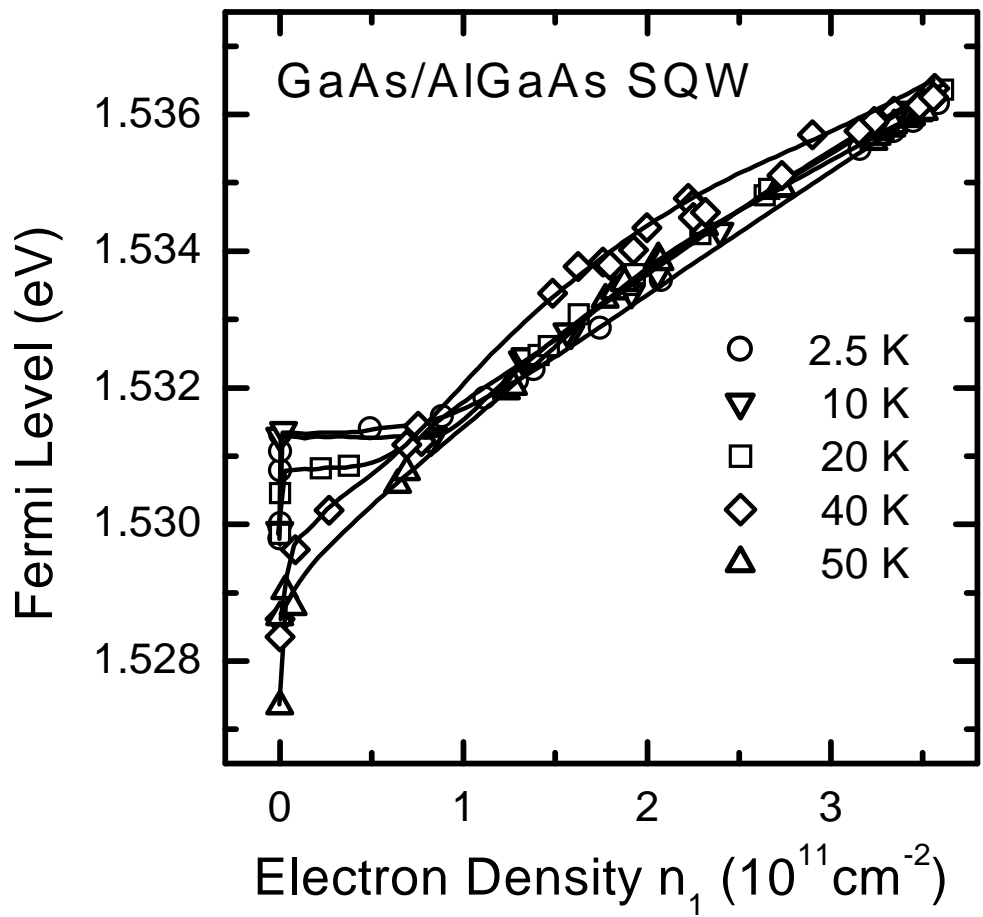
\begin{figure}[tbp]
\vspace*{1ex}
\caption{Measured isothermal curves of Fermi level versus electron density of the
first excited subband for the 2D electron gas of the modulation-doped GaAs
single quantum well. Above 30 K the discontinuity in $n_1$ is no longer observed
(see text). }
\label{LOs}
\end{figure}


\begin{references}
\bibitem{notex01a}  {\normalsize We note that the same dependence of $E_{01}$
on bias has been determined from inelastic light scattering by elementary
excitations of the 2D gas associated with electronic transitions between
subband 0 and 1. }
\end{references}
\end{document}